\documentclass[reprint,amsmath,amssymb,aps,pra,floatfix,longbibliography,superscriptaddress]{revtex4-2}
\usepackage{amsthm}

\usepackage{mathtools}
\usepackage{mathrsfs}
\usepackage{bm}
\usepackage{braket}
\usepackage{graphicx}
\usepackage{booktabs}
\usepackage{microtype}
\usepackage[
colorlinks,
linkcolor=blue,
citecolor=blue,
urlcolor=blue
]{hyperref}

\newtheorem{theorem}{Theorem}

\theoremstyle{definition}

\newcommand{\Ima}{\mathscr{I}_{g}}

\newcommand{\Tr}{\operatorname{Tr}}

\newcommand{\ketbra}[1]{\ket{#1}\!\bra{#1}}

\begin{document}

\title{Catalytic imaginarity transformations}



\author{Ying-Hui Yang}
\affiliation{School of Mathematics and Information Science,
Henan Polytechnic University, Jiaozuo 454000, China}

\author{Chen-Xi Wang}
\affiliation{School of Mathematics and Information Science,
Henan Polytechnic University, Jiaozuo 454000, China}

\author{Xiao-Li Wang}
\email{abwxl@126.com}
\affiliation{School of Mathematics and Information Science,
Henan Polytechnic University, Jiaozuo 454000, China}

\author{Shao-Ming Fei}
\affiliation{School of Mathematical Sciences,
Capital Normal University, Beijing 100048, China}

\begin{abstract}

State conversion is a central problem in quantum resource theories. In this work, we investigate catalytic state transformations within the resource theory of imaginarity.
Catalytic imaginarity transformations enable otherwise impossible state transformations using only real operations, assisted by an auxiliary system with imaginarity that is not consumed.
For deterministic catalytic conversion, we show that a maximally imaginary pure state is the unique catalyst.
This result contrasts with the resource theories of entanglement and coherence, where maximally entangled and maximally coherent catalysts cannot enable such deterministic conversions under their respective free operations.
For stochastic-approximate catalytic conversion, if we do not require catalyst recovery upon failure, any pure state with imaginarity can serve as a catalyst, and we derive the optimal success probability.
If the catalyst is required to be recovered exactly regardless of whether the catalytic state transformation succeeds or fails, the optimal success probability coincides with that without a catalyst, and thus no advantage arises from catalysis.
If exact catalyst recovery upon failure is relaxed, we find a tradeoff relation between the success probability of a single conversion and the catalyst’s residual imaginarity after transformation. When the initial state is re-prepared and the catalyst’s residual imaginarity can be reused after failure, the success probability can be further improved over $N$ (\(N \geq 2\)) rounds of catalytic conversion, and we obtain the corresponding maximum overall success probability.

\end{abstract}

\maketitle

\section{Introduction}
\label{sec:introduction}

Quantum resource theories provide a framework
for quantifying and manipulating quantum resources
under restricted operations
~\cite{Chitambar2019,Brandao2015,Gour2017,Liu2019}.
The use of complex numbers in quantum mechanics
motivated the development of the resource theory
of imaginarity, introduced by Hickey and Gour
~\cite{Hickey2018}.
In this theory, quantum states described by real density matrices in a fixed reference basis are free states, and the free operations, called real operations, are completely positive trace-preserving maps with real Kraus operators in this basis.
Measures of imaginarity were subsequently developed
~\cite{Wu2021PRA,Du2025,Guo2025,
GuoHuang2025}, and the operational roles of
imaginarity in quantum state discrimination and
imaginarity erasure were investigated
~\cite{Wu2021PRL,Shi2025}.
Further aspects of imaginarity have also been
explored, including imaginarity witnesses
~\cite{Fernandes2024}, broadcasting of imaginarity
~\cite{ZhangLi2024,ZhangLiLuo2024},
imaginarity in distributed scenarios
~\cite{Wu2024Distributed}, connections between
imaginarity and entanglement and coherence
~\cite{Sun2025,GuoGuan2026}, and
imaginarity-assisted exact simulation of unitary
operations~\cite{Nakayama2026}.

State conversion is a central problem in quantum
resource theories, encompassing deterministic
conversion and stochastic conversion~\cite{Vidal1999,Vidal2000,Regula2022,
Kondra2025}.
In the resource theory of imaginarity,
Hickey \emph{et al.} established necessary and
sufficient conditions for deterministic
transformations between pure states in a
$d$-dimensional Hilbert space under
real operations~\cite{Hickey2018}.
Wu \emph{et al.} subsequently characterized
deterministic conversions for all single-qubit states
and derived the optimal success probabilities
for stochastic conversions between pure
states~\cite{Wu2021PRL,Wu2021PRA}.
Kondra \emph{et al.} extended the analysis
to conversions from pure states to
arbitrary states, obtaining optimal
success probabilities for stochastic conversions
and stochastic-approximate conversions~\cite{Kondra2023}.
Du \emph{et al.} developed convex-roof and operational
measures of imaginarity and provided operations to realize conversions from real states to mixed states~\cite{Du2025}.

A quantum catalyst is an auxiliary state that
assists a state conversion and is recovered
exactly when the conversion succeeds.
Quantum catalysis has been investigated in
several resource theories~\cite{LipkaBartosik2024},
including entanglement~\cite{Datta2023},
coherence~\cite{Bu2016}, and quantum
thermodynamics~\cite{BrandaoThermo2015,Ng2015}.
In the resource theories of entanglement and
coherence, catalysts can both enable otherwise impossible
deterministic conversions~\cite{JonathanPlenio1999,
DaftuarKlimesh2001,Turgut2007,Bu2016}
and increase the optimal success probabilities
of stochastic conversions~\cite{Feng2004,
Feng2005,Bu2016,Fang2018,Liu2020}.
For stochastic conversions, however, requiring
the catalyst to be recovered exactly when the
conversion succeeds does not impose the same
requirement when the conversion fails.
A failed conversion may therefore change the
catalyst state, affecting its use in subsequent
rounds.
Catalyst reuse has also been studied in resource
distillation and noisy entanglement
transformations~\cite{FangLiu2026,Sharma2026}.
These considerations raise the question of
whether catalytic assistance can similarly
increase the conversion probability under real operations,
and how recovery after failure affects the
success probability and subsequent reuse.

In this work, we investigate catalytic state
conversions in the resource theory of imaginarity.
The input states and catalysts are pure, while the target states are arbitrary (either pure or mixed).
For a target state with greater imaginarity
than the input state, we find that deterministic catalytic
conversion is possible if and only if the
pure catalyst is maximally imaginary. This differs from the resource theories of entanglement and coherence. In those settings, maximally entangled and maximally coherent catalysts for entanglement and coherence cannot support such deterministic conversions under their respective free operations.
For stochastic conversions and stochastic-approximate conversions, nonmaximally imaginary catalysts
can improve the success probabilities of catalytic state conversions.
The optimal success probabilities are obtained. 
If the catalyst is required to be recovered exactly no matter whether the catalytic state transformation succeeds or fails, we find that catalytic state conversion is identical to state conversion without a catalyst. That is, the catalyst cannot improve the success probability of state conversion. This also differs from the case of coherence. If we drop the requirement for exact recovery upon failure, we find a tradeoff between the optimal success probability of the current conversion and the imaginarity retained for subsequent conversions. When the retained imaginarity is reused and fresh initial states are re-prepared, the optimal overall success probability can be improved after $N (N \geq 2)$ rounds of catalytic state conversion.

The remainder of this paper is organized as follows.
Section~\ref{sec:preliminaries} reviews the
resource theory of imaginarity and the relevant
results for state conversions without catalysts.
Section~\ref{sec:catalytic} establishes the
criterion for deterministic catalytic conversion
and derives the optimal success probabilities
for stochastic and stochastic-approximate
conversions.
Section~\ref{sec:failure_outcomes} examines how
the catalyst recovery condition after a failed
conversion affects the optimal success probability.
Section~\ref{sec:repeated_catalytic-viii}
investigates sequential conversion attempts
in which the catalyst is reused after failure,
and determines the optimal overall success probability.
Section~\ref{sec:conclusion} concludes the paper.

\section{Preliminaries}
\label{sec:preliminaries}

\subsection{Resource theory of imaginarity}
\label{subsec:imaginarity}

In the resource theory of imaginarity, suppose that an orthonormal basis $\{\ket{j}\}_{j=1}^{d}$ is the fixed reference basis for a
$d$-dimensional Hilbert space $\mathcal H$. A state is said to be free
if its density matrix is real in this reference basis. Otherwise, the state is
a resource state. A state $\rho$ is free if and only if its density matrix satisfies
\begin{equation}
    \rho=\rho^{T},
    \label{eq:real-state}
\end{equation}
where $T$ denotes transposition in the reference basis.
The free operations are real operations
\cite{Hickey2018}. 

A trace-preserving quantum operation $\Lambda$
admits a Kraus representation of the form
\begin{equation}
    \Lambda(\rho)=\sum_j K_j\rho K_j^\dagger,
\end{equation}
where $\sum_j K_j^\dagger K_j=\mathbb{I}$ and $\mathbb{I}$ is the identity operator.
The operation $\Lambda$ is real~\cite{Hickey2018} if it admits
a Kraus representation in which
$\langle m|K_j|n\rangle\in\mathbb{R}$ in the fixed reference basis for all $j,m,n$.

To quantify imaginarity, Wu \emph{et al.} defined the geometric measure of imaginarity~\cite{Wu2021PRA}. For a pure state $\ket{\psi}$, its geometric imaginarity is
\begin{equation}
    \Ima(\ket{\psi})
    =
    1-
    \max_{\ket{\phi}\in\mathcal R}
    \left|\langle\phi|\psi\rangle\right|^2,
    \label{eq:Ig-pure-definition}
\end{equation}
where $\mathcal R$ is the set of real pure states. Applying the convex roof, the geometric imaginarity of a mixed state \(\rho\) is given by
\begin{equation}
    \Ima(\rho)
    =
    \min_{\{p_j,\ket{\psi_j}\}}
    \sum_j p_j\Ima(\ket{\psi_j}),
    \label{eq:Ig-convex-roof}
\end{equation}
where $\rho=\sum_j p_j\ketbra{\psi_j}$ and the minimum is taken over all pure state decompositions of $\rho$.

The geometric imaginarity of a pure state $\ket{\psi}$ can be calculated by
\begin{equation}
    \Ima(\ket{\psi})
    =
    \frac{
        1-\left|\langle\psi^*|\psi\rangle\right|
    }{2},
    \label{eq:Ig-pure-closed}
\end{equation}
where $\ket{\psi^*}$ is the conjugate complex of $\ket{\psi}$ in the reference basis. Kondra \emph{et al.} generalized the analytical expression to an
arbitrary quantum state $\rho$~\cite{Kondra2023}, giving
\begin{equation}
    \Ima(\rho)
    =
    \frac{1-\sqrt{F(\rho,\rho^{T})}}{2},
    \label{eq:Ig-closed}
\end{equation}
where
\begin{equation}
    F(\rho,\sigma)
    =
    \left[
    \operatorname{Tr}
    \sqrt{\sqrt{\rho}\sigma\sqrt{\rho}}
    \right]^2
    \label{eq:fidelity}
\end{equation}
is the squared Uhlmann fidelity \cite{Jozsa1994}.

Geometric imaginarity is convex and strongly
monotonic under real operations~\cite{Wu2021PRA}.
For any ensemble $\{p_j,\rho_j\}$ of density
operators, convexity gives
\[
    \Ima\left(\sum_j p_j\rho_j\right)
    \leq \sum_j p_j\Ima(\rho_j).
\]
For a real operation with Kraus operators
$\{K_j\}$, strong monotonicity gives
\begin{equation}
    \Ima(\rho)
    \geq
    \sum_j p_j\Ima(\rho_j),
    \label{eq:strong-monotonicity}
\end{equation}
where $p_j=\operatorname{Tr}(K_j\rho K_j^\dagger)$
and $\rho_j=K_j\rho K_j^\dagger/p_j$
for $p_j>0$.

For arbitrary quantum states $\rho$ and $\tau$,
we can derive the following tensor product relation
for geometric imaginarity applying
Eq.~\eqref{eq:Ig-closed} and the multiplicativity
of fidelity,
\begin{equation}
    \Ima(\rho\otimes\tau)
    =
    \Ima(\rho)
    +
    \Ima(\tau)
    -
    2\Ima(\rho)\Ima(\tau).
    \label{eq:Ig-product}
\end{equation}

\subsection{Single-copy state transformations}
\label{subsec:single-copy}

A fundamental problem of state conversion in the resource theory of imaginarity is whether a given state \(\rho\) can be transformed into a target state \(\sigma\) via real quantum operations. A deterministic conversion $\rho \to \sigma$ is possible if there exists a trace-preserving real operation $\Lambda$ satisfying
\begin{equation}
    \sigma=\Lambda(\rho).
    \label{eq:deterministic-conversion}
\end{equation}
The deterministic conversion for a pure state transformation
$\ket{\psi}\to\ket{\phi}$ under real operations
is possible if and only if~\cite{Hickey2018,Wu2021PRA}
\begin{equation}
    \Ima(\ket{\psi})\geq\Ima(\ket{\phi}).
    \label{eq:pure-deterministic-criterion}
\end{equation}
This condition adopts the geometric imaginarity of quantum states as the criterion governing deterministic conversion between pure states.

Eq.~\eqref{eq:pure-deterministic-criterion} implies that deterministic conversion between pure states is not always possible. This motivates the study of stochastic conversion, in which the target state can be obtained probabilistically. A stochastic real operation is characterized by real Kraus operators $\{K_j\}_{j\in\mathcal S}$ satisfying

\begin{equation}
    \mathcal E_{\mathrm s}(\ketbra{\psi})
    =
    \sum_{j\in\mathcal S}K_j\ketbra{\psi}K_j^\dagger,
    \quad
    \sum_{j\in\mathcal S}K_j^\dagger K_j\leq\mathbb I.
    \label{eq:stochastic-real-operation}
\end{equation}
A stochastic conversion $\ket{\psi} \to \rho$ is achieved with probability \(p\) if there exists a stochastic real operation \(\mathcal E_{\mathrm s}\) such that

\begin{equation}
    \mathcal E_{\mathrm s}(\ketbra{\psi})=p\rho,
    \label{eq:exact-stochastic-process}
\end{equation}
where the success probability $p=\Tr\mathcal E_{\mathrm s}(\ketbra{\psi})$. When $p=1$, the conversion succeeds with certainty for input state $\ket{\psi}$. Since different admissible operations may yield different success probabilities, the optimal success probability is defined as the supremum of $p$ over all stochastic real operations.
For an arbitrary target state $\rho$ with
$\Ima(\rho)>0$, Kondra \emph{et al.} showed that the optimal success
probability of stochastic conversion is \cite{Kondra2023}
\begin{equation}
    P(\ket{\psi}\rightarrow\rho)
    =
    \min\left\{
        \frac{\Ima(\ket{\psi})}{\Ima(\rho)},
        1
    \right\}.
    \label{eq:exact-single-copy}
\end{equation}
If $\Ima(\rho)=0$, the target state is free and can be prepared
deterministically by a real operation.

If the output state \(\sigma\) is allowed to differ from the target state \(\rho\) as long as \(F(\sigma, \rho) \ge f\), the transformation is said to be approximate~\cite{Kondra2023}, where \(f \in [0, 1]\) is the prescribed fidelity threshold. For a pure input state $\ket{\psi}$, a stochastic-approximate
conversion with success probability $p$ is described by
a trace-nonincreasing real operation $\mathcal E_{\mathrm s}$
satisfying
\begin{equation}
    \mathcal E_{\mathrm s}(\ketbra{\psi})=p\sigma,
    \qquad
    F(\sigma,\rho)\geq f.
    \label{eq:stochastic-approximate-process}
\end{equation}
The optimal success probability is obtained by maximizing
$p$ over all such real operations and output
states $\sigma$,
\begin{equation}
    P_f(\ket{\psi}\rightarrow\rho)
    =
    \max_{\mathcal E_{\mathrm s},\,\sigma}
    \left\{
        p:
        \mathcal E_{\mathrm s}(\ketbra{\psi})=p\sigma,\ 
        F(\sigma,\rho)\geq f
    \right\}.
    \label{eq:pf-definition}
\end{equation}

The fidelity requirement generally allows more than one successful output
state. For a pure input state $\ket{\psi}$, Eq.~\eqref{eq:exact-single-copy} shows that the 
conversion probability to an admissible output $\sigma$ is determined by
$\Ima(\sigma)$. Maximizing the success probability therefore reduces to
finding the smallest geometric imaginarity among all states that satisfy the
fidelity requirement \cite{Kondra2023}. The minimum geometric imaginarity is denoted by \(b_f\), i.e.,
\begin{equation}
    b_f
    =
    \min_{\sigma:\,F(\sigma,\rho)\geq f}
    \Ima(\sigma).
    \label{eq:bf-definition}
\end{equation}
Kondra \emph{et al.}~\cite{Kondra2023} obtained the analytical expression
\begin{equation}
    b_f
    =
    \sin^2
    \left[
        \max
        \left\{
            \arcsin\sqrt{\Ima(\rho)}
            -
            \arccos\sqrt f,
            0
        \right\}
    \right].
    \label{eq:bf-closed}
\end{equation}
For an input state \(\ket{\psi}\), the optimal stochastic-approximate conversion
probability is then 
\begin{equation}
    P_f(\ket{\psi}\rightarrow\rho)
    =
    \begin{cases}
        1,
        & b_f=0,\\[1mm]
        \displaystyle
        \min\left\{
            \frac{\Ima(\ket{\psi})}{b_f},
            1
        \right\},
        & b_f>0.
    \end{cases}
    \label{eq:stochastic-approximate-probability}
\end{equation}
When $\Ima(\ket{\psi}) \geq b_f$, the probability reaches unity, and the conversion becomes deterministic.
When $f=1$, the output state is the target state, so
$b_1=\Ima(\rho)$. 
Eq.~\eqref{eq:stochastic-approximate-probability} then reduces to
Eq.~\eqref{eq:exact-single-copy}.

\section{Catalytic state transformations}
\label{sec:catalytic}

The optimal success probability of state conversion without catalytic assistance has been established. A natural question is whether introducing an auxiliary state as a catalyst can improve the optimal success probability. In this section, we focus on this question. 

A catalyst is an auxiliary resource state that participates in the joint operation and undergoes exact recovery when the target state is achieved.
Consider state transformations of a system
from an initial pure state $\ket{\psi}$ to
an arbitrary target state $\rho$ (whether pure or mixed), assisted
by a pure catalyst $\ket{\mu}$ in an
auxiliary system. Note that when we consider catalytic state transformations $\ket{\psi} \to \rho$, we only focus on
\begin{equation}
    \Ima(\ket{\psi}) < \Ima(\rho),
    \label{eq:nontrivial-regime}
\end{equation}
since, in this regime, deterministic conversion is impossible without a catalyst. In addition, the imaginarity of the catalyst $c=\Ima(\ket{\mu})$ is required to satisfy
\begin{equation}
    0 < c \leq \frac{1}{2},
    \label{eq:catalyst-range}
\end{equation}
since real states have zero geometric imaginarity and the maximally imaginary state has an imaginarity of $1/2$~\cite{Wu2021PRA}.

\begin{figure}[t]  
\centering  
\includegraphics[width=\columnwidth]  
{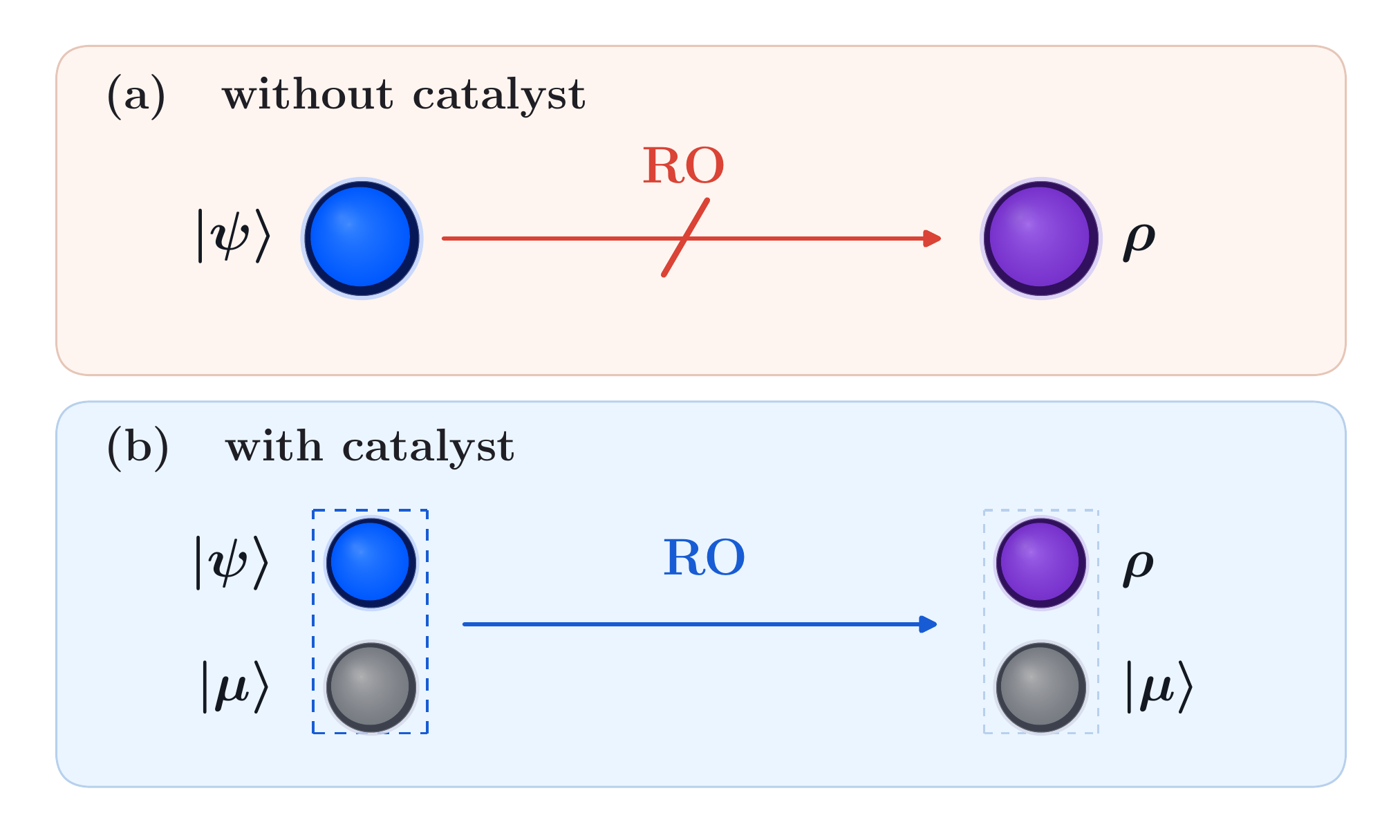}  
\caption{Schematic of deterministic catalytic conversion  
under real operations (RO).  
(a) The transformation 
$\ket{\psi}\to\rho$ cannot  
be achieved deterministically without a catalyst.  
(b) An appropriate pure catalyst $|\mu\rangle$ enables the transformation while being recovered exactly.}  
\label{fig:deterministic_catalysis}
\end{figure}

As illustrated in Fig.~\ref{fig:deterministic_catalysis}, a deterministic catalytic conversion is implemented by a real quantum operation $\mathcal{E}$ acting jointly on the input state $\ket{\psi}$ and the catalyst $\ket{\mu}$, such that
\begin{equation}    
\mathcal E\!\left(        
\ket{\psi}\!\bra{\psi}\otimes\ket{\mu}\!\bra{\mu}    
\right)    
=    
\rho\otimes\ket{\mu}\!\bra{\mu}.    
\label{eq:deterministic-catalytic-operation}
\end{equation}

\begin{theorem}
\label{thm:maximal-catalyst}
The deterministic catalytic transformation in
Eq.~\eqref{eq:deterministic-catalytic-operation} is possible if and only if the catalyst $\mu$ is a maximally imaginary pure state, i.e.,
$c=\Ima(\ket{\mu})=\frac{1}{2}$.
\end{theorem}

\begin{proof}
\emph{Sufficiency.}
For a catalyst $\ket{\mu}$ with the imaginarity $c=1/2$,
the tensor product relation in
Eq.~\eqref{eq:Ig-product} gives
\[
    \Ima(\ket{\psi}\otimes\ket{\mu})
    =\frac{1}{2}
    =\Ima(\rho\otimes\ketbra{\mu}).
\]
Then the optimal success probability of stochastic conversion in Eq.~\eqref{eq:exact-single-copy} yields
\[
    P\!\left(
        \ket{\psi}\otimes\ket{\mu}
        \rightarrow \rho\otimes\ketbra{\mu}
    \right)=1.
\]
Thus, the deterministic catalytic conversion
in Eq.~\eqref{eq:deterministic-catalytic-operation}
is possible.

\emph{Necessity.}
Suppose the deterministic catalytic conversion
in Eq.~\eqref{eq:deterministic-catalytic-operation}
is possible. The joint input and output states satisfy
\[
    \Ima(\ket{\psi}\otimes\ket{\mu})
    \geq \Ima(\rho\otimes\ketbra{\mu}),
\]
by monotonicity of geometric imaginarity under
real operations. It follows
\begin{align}
    0 &\geq
    \Ima(\rho\otimes\ketbra{\mu})
    -\Ima(\ket{\psi}\otimes\ket{\mu})
    \nonumber\\
    &=(1-2c)
    \bigl[\Ima(\rho)-\Ima(\ket{\psi})\bigr],
    \label{eq:det_cat_gap}
\end{align}
where the last equality holds by the tensor product relation in Eq.~\eqref{eq:Ig-product}.
Since $\Ima(\ket{\psi})<\Ima(\rho)$,
Eq.~\eqref{eq:det_cat_gap} requires $c=1/2$.
\end{proof}

Theorem~\ref{thm:maximal-catalyst} shows that a maximally imaginary pure state is the unique catalyst to realize deterministic catalytic transformations. It differs from entanglement and coherence. For entanglement and coherence, deterministic catalysis of pure states is governed by majorization. Under LOCC (incoherent operations), 
the maximally entangled state (coherent) cannot serve as a catalyst to realize deterministic catalytic transformations~\cite{JonathanPlenio1999,Bu2016}.

A related result was recently obtained by
Nakayama \emph{et al.}~\cite{Nakayama2026},
who studied the exact simulation of unitary
operations assisted by imaginarity.
They showed that, under real orthogonal operations,
only maximally imaginary resource states can enable
the implementation of non-real unitary gates,
and such states are sufficient to simulate arbitrary
unitary operations.
Although their work focuses on unitary simulation,
whereas ours considers catalytic state transformations,
both results highlight the special role of maximal
imaginarity in overcoming the limitations imposed
by real operations.


Although deterministic conversion cannot be realized with a nonmaximally imaginary catalyst, such a catalyst can still increase the optimal success probability of stochastic conversion. A stochastic catalytic
conversion $\ket{\psi} \to \rho$, with success probability $p$, is realized
by a trace-nonincreasing real operation $\mathcal E_{\rm s}$ satisfying
\begin{equation}
    \mathcal E_{\rm s}\!\left(
        \ket{\psi}\!\bra{\psi}\otimes\ket{\mu}\!\bra{\mu}
    \right)
    =
    p\,\rho\otimes\ket{\mu}\!\bra{\mu}.
    \label{eq:probabilistic-catalytic-operation}
\end{equation}
We now characterize the optimal success probability of such stochastic catalytic conversions.

\begin{theorem}
\label{thm:exact-catalytic}
The optimal success probability for stochastic
catalytic conversion is
\begin{equation}
    P^{\rm cat}(\ket{\psi}\to\rho;\ket{\mu})
    =
    \frac{
        c+(1-2c)\Ima(\ket{\psi})
    }{
        c+(1-2c)\Ima(\rho)
    }.
    \label{eq:exact-catalytic-probability}
\end{equation}
Moreover, catalytic assistance strictly increases
the optimal success probability,
\begin{equation}
    P^{\rm cat}(\ket{\psi}\to\rho;\ket{\mu})
    >P(\ket{\psi}\to\rho).
    \label{eq:exact-catalytic-advantage}
\end{equation}
\end{theorem}

\begin{proof}
By the tensor-product relation in
Eq.~\eqref{eq:Ig-product}, we have
\[
\begin{aligned}
    \Ima(\ket{\psi}\otimes\ket{\mu})
    &=c+(1-2c)\Ima(\ket{\psi}),\\
    \Ima(\rho\otimes\ketbra{\mu})
    &=c+(1-2c)\Ima(\rho).
\end{aligned}
\]
It follows that
\[
    \Ima(\ket{\psi}\otimes\ket{\mu})
    \leq \Ima(\rho\otimes\ketbra{\mu}),
\]
since $\Ima(\ket{\psi})<\Ima(\rho)$.
The optimal success probability formula for 
stochastic conversion in
Eq.~\eqref{eq:exact-single-copy} therefore gives
\begin{equation}
\begin{aligned}
    P^{\rm cat}(\ket{\psi}\to\rho;\ket{\mu})
    =
    \frac{
        c+(1-2c)\Ima(\ket{\psi})
    }{
        c+(1-2c)\Ima(\rho)
    }.
\end{aligned}
    \label{eq:joint-catalytic-probability}
\end{equation}

For the conversion without a catalyst,
Eq.~\eqref{eq:exact-single-copy} gives
$P(\ket{\psi}\to\rho)
=\Ima(\ket{\psi})/\Ima(\rho)$.
Comparing the two optimal success probabilities,
we find
\begin{align}
    &P^{\rm cat}(\ket{\psi}\to\rho;\ket{\mu})
    -P(\ket{\psi}\to\rho)
    \nonumber\\
    &\qquad=
    \frac{
        c[\Ima(\rho)-\Ima(\ket{\psi})]
    }{
        \Ima(\rho)
        [c+(1-2c)\Ima(\rho)]
    }
    >0,
    \label{eq:exact-catalytic-gap}
\end{align}
since
$0<c \leq \frac{1}{2}$ and $\Ima(\ket{\psi})<\Ima(\rho)$.
This proves Eq.~\eqref{eq:exact-catalytic-advantage}.
\end{proof}

Theorem~\ref{thm:exact-catalytic} shows that every nonmaximally imaginary pure catalyst strictly increases the optimal success probability
compared with that without catalytic assistance. Moreover, the optimal probability depends only on the geometric imaginarity \(c\) of the catalyst \(\ket{\mu}\), rather than on the specific state. Two pure states with the same geometric imaginarity \(c\) can be interconverted via real orthogonal transformations~\cite{Wu2021PRA}. Therefore, any state with geometric imaginarity \(c\) can be used in the catalytic conversion and recovered exactly. When \(c=1/2\), the optimal probability in Theorem~\ref{thm:exact-catalytic} reaches unity. In this case, the stochastic conversion becomes the deterministic conversion described in Theorem~\ref{thm:maximal-catalyst}.

In a stochastic-approximate catalytic conversion, the input state \(\ket{\psi}\) is transformed into a state \(\sigma\) satisfying \(F(\sigma,\rho)\geq f\) with probability \(p\), while the catalyst is recovered exactly. The conversion is realized by a real operation \(\mathcal E_{\rm s}\) that yields success with probability \(p\):
\begin{equation}
    \mathcal E_{\rm s}\!\left(
        \ketbra{\psi}\otimes\ketbra{\mu}
    \right)
    =
    p\,\sigma\otimes\ketbra{\mu},
    \quad
    F(\sigma,\rho)\geq f.
    \label{eq:approx-catalytic-operation}
\end{equation}

Recall that $b_f$, defined in
Eq.~\eqref{eq:bf-definition}, is the minimum
geometric imaginarity among states satisfying
the fidelity requirement. Within the nontrivial regime where catalytic assistance is required, the stochastic
catalytic conversion condition \(\Ima(\ket{\psi}) < \Ima(\rho)\) naturally relaxes to \(\Ima(\ket{\psi}) < b_f\) for stochastic-approximate catalytic conversion.
The optimal success probability for
stochastic-approximate catalytic conversion is
then obtained by maximizing the exact catalytic
conversion probability over all admissible states
$\sigma$, i.e.,
\begin{equation}
    P_f^{\rm cat}(\ket{\psi}\to\rho;\ket{\mu})=
    \max_{\sigma:\,F(\sigma,\rho)\geq f}
    P\!\left(
        \ket{\psi}\otimes\ket{\mu}
        \to \sigma\otimes\ketbra{\mu}
    \right).
    \label{eq:cat-prob-f}
\end{equation}

\begin{theorem}
\label{thm:approx-catalytic}
The optimal success probability for stochastic-approximate catalytic conversion is
\begin{equation}
    P_f^{\rm cat}(\ket{\psi}\to\rho;\ket{\mu})
    =
    \frac{
        c+(1-2c)\Ima(\ket{\psi})
    }{
        c+(1-2c)b_f
    },
    \label{eq:approx-catalytic-probability}
\end{equation}
Moreover, catalytic assistance strictly increases
the optimal success probability, i.e.,
\begin{equation}
    P_f^{\rm cat}(\ket{\psi}\to\rho;\ket{\mu})
    >P_f(\ket{\psi}\to\rho).
    \label{eq:approx-catalytic-advantage}
\end{equation}
\end{theorem}

\begin{proof}
By the definition of $b_f$ in
Eq.~\eqref{eq:bf-definition}, every state $\sigma$
satisfying $F(\sigma,\rho)\geq f$ obeys
\[
    \Ima(\sigma)\geq b_f>\Ima(\ket{\psi}).
\]
Theorem~\ref{thm:exact-catalytic} therefore gives
the optimal success probability for exact catalytic
conversion to $\sigma$:
\begin{equation}
    P^{\rm cat}(\ket{\psi}\to\sigma;\ket{\mu})
    =
    \frac{
        c+(1-2c)\Ima(\ket{\psi})
    }{
        c+(1-2c)\Ima(\sigma)
    }.
\end{equation}
Since $1-2c\geq0$, the optimal success probability
$P^{\rm cat}(\ket{\psi}\to\sigma;\ket{\mu})$
is nonincreasing as $\Ima(\sigma)$ increases. The maximum in
Eq.~\eqref{eq:cat-prob-f} is therefore attained
at $\Ima(\sigma)=b_f$, giving
Eq.~\eqref{eq:approx-catalytic-probability}.

Since $\Ima(\ket{\psi})<b_f$,
Eq.~\eqref{eq:stochastic-approximate-probability}
gives the optimal success probability for
stochastic-approximate conversion without
catalytic assistance:
\[
    P_f(\ket{\psi}\to\rho)
    =
    \frac{\Ima(\ket{\psi})}{b_f}.
\]
Subtracting this probability from
Eq.~\eqref{eq:approx-catalytic-probability},
one can obtain
\begin{align}
    &P_f^{\rm cat}(\ket{\psi}\to\rho;\ket{\mu})
    -P_f(\ket{\psi}\to\rho)
    \nonumber\\
    &\qquad=
    \frac{
        c[b_f-\Ima(\ket{\psi})]
    }{
        b_f[c+(1-2c)b_f]
    }>0.
\end{align}
This completes the proof.
\end{proof}

\begin{figure}[t]
    \centering
    \includegraphics[width=0.95\columnwidth]
    {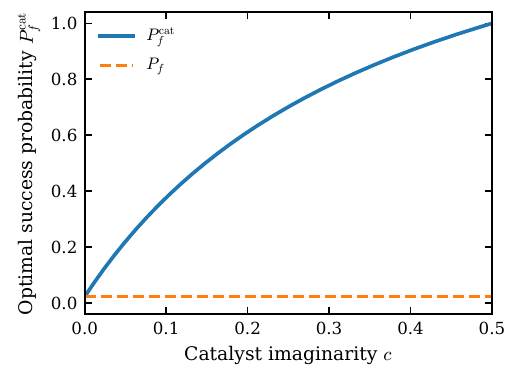}
    \caption{
    Optimal success probability $P_f^{\rm cat}$
    for $\Ima(\ket{\psi})=0.005$ and $b_f=0.22$,
    as the imaginarity $c$ of the catalyst
    varies from $0$ to $1/2$.
    The dashed horizontal line indicates the success probability without catalytic assistance
    $P_f(\ket{\psi}\to\rho)=
    \Ima(\ket{\psi})/b_f$.
    }
    \label{fig:approx-catalytic-advantage}
\end{figure}

Theorem~\ref{thm:approx-catalytic} unifies several of the preceding results as special cases. If $f=1$, then
$b_f=\Ima(\rho)$. Eq.~\eqref{eq:approx-catalytic-probability} then
reduces to the stochastic
catalytic conversion probability in Eq.~\eqref{eq:exact-catalytic-probability}. If $c=0$,
Eq.~\eqref{eq:approx-catalytic-probability} reduces to the optimal success probability
without catalytic assistance in Eq.~\eqref{eq:stochastic-approximate-probability}.
If $c=1/2$, Eq.~\eqref{eq:approx-catalytic-probability} gives unit
success probability, recovering
Theorem~\ref{thm:maximal-catalyst}.

Differentiating Eq.~\eqref{eq:approx-catalytic-probability}
with respect to the imaginarity $c$ of the
catalyst yields
\begin{align}
\frac{dP_f^{\mathrm{cat}}}{dc}
&=
\frac{
b_f-\Ima(\ket{\psi})
}{
\left[c+(1-2c)b_f\right]^2
}
>0,
\\
\frac{d^2P_f^{\mathrm{cat}}}{dc^2}
&=
-
\frac{
2(1-2b_f)\left[b_f-\Ima(\ket{\psi})\right]
}{
\left[c+(1-2c)b_f\right]^3
}
\leq 0.
\label{eq:diminishing_return}
\end{align}

The dependence of the optimal success
probability on $c$ is illustrated in
Fig.~\ref{fig:approx-catalytic-advantage}.
The positive first derivative means that a larger imaginarity $c$ of the catalyst $\ket{\mu}$ always increases the success probability, while the nonpositive second derivative means that this growth slows down as $c$ gets larger. In other words, each additional increase in the imaginarity of the catalyst yields a smaller marginal gain, a classic signature of diminishing returns.

\section{Catalyst recovery and resource retention}
\label{sec:failure_outcomes}
\label{sec:catalyst-recovery}



The stochastic-approximate catalytic conversions studied in Sec.~\ref{sec:catalytic} improve the success probability, and the catalyst is recovered exactly when the conversion succeeds.
However, the preceding analysis does not
consider how much imaginarity the catalyst
retains when the conversion fails.
In probabilistic coherence distillation,
Fang \emph{et al.}~\cite{Fang2018}
considered exact catalyst recovery after both
success and failure.
They showed through an explicit example under
dephasing covariant incoherent operations that
catalytic assistance can still improve the
distillation success probability.
This raises the question of whether the
catalytic advantage established in
Sec.~\ref{sec:catalytic} persists under
the same recovery requirement.

In a stochastic-approximate catalytic conversion,
the success branch corresponds to obtaining
$\sigma\otimes\ketbra{\mu}$ with
$F(\sigma,\rho)\geq f$, while the failure branch
corresponds to an unsuccessful conversion.
 For a maximally imaginary
catalyst, the optimal success probability
is one by Theorem~\ref{thm:maximal-catalyst}. Thus, only $0<c<1/2$ needs to be considered when considering the failure branch.

Let $\mathcal E_{\rm succ}$ and
$\mathcal E_{\rm fail}$ denote the corresponding
trace-nonincreasing real operations, with
$\mathcal E_{\rm succ}+\mathcal E_{\rm fail}$
being trace preserving. If the catalyst is also
required to be recovered exactly after failure,
these operations satisfy
\begin{align}
    \mathcal E_{\rm succ}\!\left(
        \ketbra{\psi}\otimes\ketbra{\mu}
    \right)
    &=
    p\,\sigma\otimes\ketbra{\mu},
    \quad F(\sigma,\rho)\geq f,
    \label{eq:exact-recovery-success}\\
    \mathcal E_{\rm fail}\!\left(
        \ketbra{\psi}\otimes\ketbra{\mu}
    \right)
    &=
    (1-p)\,\tau\otimes\ketbra{\mu},
    \label{eq:exact-recovery-failure}
\end{align}
where $\tau$ is the state of the
system after failure, i.e., $F(\tau,\rho)< f$.

The following theorem answers the question above
in the negative.

\begin{theorem}
\label{thm:all-outcome-exact-return}
When exact catalyst recovery is required on both the success and failure
branches, the nonmaximally imaginary catalyst cannot enhance the optimal success probability, i.e.,
\begin{equation}
    P^{\rm cat}_{f,\mathrm{rec}}(\ket{\psi}\to\rho;\ket{\mu})
    =
    \frac{\Ima(\ket{\psi})}{b_f}
    =
    P_f(\ket{\psi}\to\rho).
    \label{eq:IV_exact_return_result}
\end{equation}
\end{theorem}

\begin{proof}
For any state $\sigma$ satisfying $F(\sigma,\rho)\ge f$,
Eq.~\eqref{eq:bf-definition} implies $\Ima(\sigma)\ge b_f$.
The tensor product relation in Eq.~\eqref{eq:Ig-product} yields
\begin{equation}
    \Ima(\ket{\psi}\otimes\ket{\mu})
    =c+(1-2c)\Ima(\ket{\psi}).
    \label{eq:input-imag}
\end{equation}

Similarly, the success and failure outputs satisfy
\[
\begin{aligned}
    \Ima(\sigma\otimes\ketbra{\mu})
    &\ge c+(1-2c)b_f,\\
    \Ima(\tau\otimes\ketbra{\mu})
    &\ge c,
\end{aligned}
\]
where the second bound follows from $\Ima(\tau)\ge 0$.
By strong monotonicity and convexity of geometric
imaginarity under real operations, we obtain
\[
\begin{aligned}
\Ima(\ket{\psi}\otimes\ket{\mu})
&\ge p\,\Ima(\sigma\otimes\ketbra{\mu})\\
&\quad +(1-p)\Ima(\tau\otimes\ketbra{\mu})\\
&\ge p[c+(1-2c)b_f]+(1-p)c.
\end{aligned}
\]
It follows from Eq.~\eqref{eq:input-imag} that
\[
    c+(1-2c)\Ima(\ket{\psi})
    \ge c+p(1-2c)b_f.
\]
Thus
\begin{equation}
    p\le \frac{\Ima(\ket{\psi})}{b_f}.
    \label{eq:IV_exact_balance}
\end{equation}
This upper bound equals the optimal stochastic-approximate conversion probability
without catalytic assistance.

Finally, the bound is attainable.
Let $\{\mathcal F_{\rm succ},\mathcal F_{\rm fail}\}$
be an optimal real operation without a catalyst.
The operations
\[
    \mathcal E_{\rm succ}
    =\mathcal F_{\rm succ}\otimes \mathbb{I},
    \qquad
    \mathcal E_{\rm fail}
    =\mathcal F_{\rm fail}\otimes \mathbb{I}
\]
are real and leave the catalyst unchanged in both branches.
$P_f(\ket{\psi}\to\rho)$ is achieved, proving
Eq.~\eqref{eq:IV_exact_return_result}.
\end{proof}

Theorem~\ref{thm:all-outcome-exact-return}
shows that requiring exact catalyst recovery
in both branches leaves the optimal success
probability equal to that without catalytic
assistance. We next consider how this probability
changes when the requirement for exact catalyst recovery after
failure is relaxed.

On the failure branch, suppose the catalyst $\ket{\mu}$
is left in a pure state $\ket{\mu'}$ with
retained imaginarity $c'=\Ima(\ket{\mu'})$,
where $0\leq c'\leq c$.
The failure output is then
\begin{equation}
    \mathcal E_{\rm fail}\!\left(
        \ketbra{\psi}\otimes\ketbra{\mu}
    \right)
    =
    (1-p)\tau\otimes\ketbra{\mu'}.
    \label{eq:failure-output}
\end{equation}

\begin{theorem}
    \label{thm:failure_imaginarity}
For a pure catalyst with initial imaginarity
$c$ and retained imaginarity $c'$ after failure, the optimal success probability for stochastic-approximate catalytic conversion is
\begin{equation}
P_f^\star(c,c')
=
\frac{
c-c'+(1-2c)\Ima(\ket{\psi})
}{
c-c'+(1-2c)b_f
}.
\label{eq:recovery_tradeoff}
\end{equation}
Moreover, a strict catalytic advantage is
obtained if and only if $c'<c$:
\begin{equation}
P_f^\star(c,c')
>
P_f(\ket{\psi}\to\rho)
\quad\Longleftrightarrow\quad
c'<c.
\label{eq:recovery_advantage}
\end{equation}
\end{theorem}

The proof is given in Appendix~\ref{app:retained_imaginarity}.

\begin{figure}[t]
    \includegraphics[width=0.95\columnwidth]{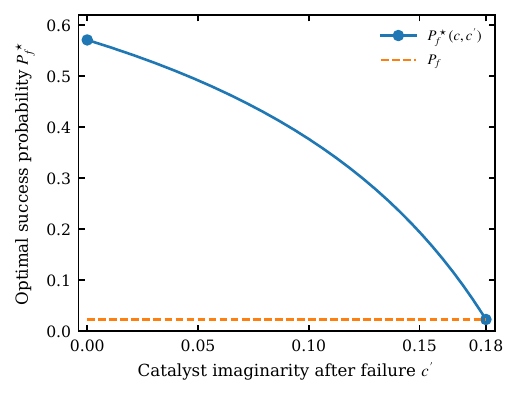}
    \caption{
    Optimal success probability $P_f^\star(c,c')$ for
$\Ima(\ket{\psi})=0.005$, $c=0.18$, and $b_f=0.22$,
as the imaginarity of the catalyst retained after failure varies from
$c'=0$ to $c'=c$.
The dashed horizontal line indicates the unassisted probability \(P_f(\ket{\psi}\to\rho)=\Ima(\ket{\psi})/b_f\).
    }
    \label{fig:recovery-tradeoff}
\end{figure}

With the initial imaginarity of the catalyst
fixed at $c$, Eq.~\eqref{eq:recovery_tradeoff}
relates the optimal success probability of
the current conversion to the imaginarity
$c'$ retained after failure.
Differentiating with respect to $c'$ yields
\begin{equation}
    \frac{\partial P_f^\star(c,c')}{\partial c'}
    =
    -\frac{
        (1-2c)[b_f-\Ima(\ket{\psi})]
    }{
        [c-c'+(1-2c)b_f]^2
    }
    <0.
    \label{eq:recovery_derivative}
\end{equation}
Thus, for fixed $c$, the optimal success
probability strictly decreases with $c'$,
as illustrated in Fig.~\ref{fig:recovery-tradeoff}.

At $c'=c$, all the imaginarity of the catalyst is
retained after failure, and
$P_f^\star(c,c)=P_f(\ket{\psi}\to\rho)$,
recovering Theorem~\ref{thm:all-outcome-exact-return}.
At $c'=0$, no imaginarity is retained after
failure, and Eq.~\eqref{eq:recovery_tradeoff}
reduces to Theorem~\ref{thm:approx-catalytic},
giving the maximum success probability. For $0<c'<c$, part of the imaginarity
of the catalyst is retained, while
a strict catalytic advantage remains,
as shown by Eq.~\eqref{eq:recovery_advantage}.

\section{Repeated catalytic state conversions}
\label{sec:repeated_catalytic-viii}


A failed stochastic-approximate catalytic conversion
may leave the catalyst in a pure state $\ket{\mu'}$
with nonzero imaginarity.
If fresh copies of $\ket{\psi}$ are available,
the imaginarity retained in $\ket{\mu'}$
can be reused in subsequent attempts to obtain
the target state $\rho$. In this section, we investigate whether such
reuse can increase the overall success probability
and how the optimal probability varies with the
number of conversion rounds.


We first consider two successive rounds of
stochastic-approximate catalytic conversion.
 The first round starts from the joint state $\ket{\psi}\otimes\ket{\mu_0}$, where $\ket{\mu_0}$ is the initial catalyst with geometric imaginarity $c_0$. If the first round fails, the second round takes the joint state $\ket{\psi} \otimes \ket{\mu_1}$ as input, where $\ket{\mu_1}$ is the remaining pure catalyst with imaginarity $c_1$.
Since the catalyst is not needed after the
second round, we take $c_2=0$.



Let $a=\Ima(\ket{\psi})$ and $b=b_f$,
with $a<b$.
From Theorem~\ref{thm:failure_imaginarity},
the maximum success probability is
\begin{equation}
    p^\star(c,c')
    =
    \frac{
        c-c'+(1-2c)a
    }{
        c-c'+(1-2c)b
    },
    \label{eq:v-two-success}
\end{equation}
and the corresponding failure probability is
\begin{equation}
    q^\star(c,c')
    =
    1-p^\star(c,c')
    =
    \frac{
        (1-2c)(b-a)
    }{
        c-c'+(1-2c)b
    }.
    \label{eq:v-two-failure}
\end{equation}

Since the second round is performed only if
the first fails, the overall success
probability is
\begin{equation}
    P_2(c_1)
    =
    1-q^\star(c_0,c_1)q^\star(c_1,0).
    \label{eq:v-two-total-success}
\end{equation}
Maximizing $P_2(c_1)$ over
$0\leq c_1\leq c_0$ gives the optimal overall success probability
\begin{equation}
    P_2^\star
    =
    P_2(c_1^\star)
    =
    1-
    \left[
        \frac{2(b-a)}
        {(1-2c_0)^{-1/2}-(1-2b)}
    \right]^2,
    \label{eq:v-two-optimal-probability}
\end{equation}
at retained imaginarity
\begin{equation}
    c_1^\star
    =
    \frac{1-\sqrt{1-2c_0}}{2}.
    \label{eq:v-two-optimal-c1}
\end{equation}
Note that the retained imaginarity $c_1^\star$
depends only on $c_0$ and satisfies
$p^\star(c_0,c_1^\star)=p^\star(c_1^\star,0)$.
The derivation is given in
Appendix~\ref{app:repeated-recovery-proof}.

\begin{table*}[t]
\caption{\label{tab:v-two-attempts}
Comparison of two-round conversion strategies
for $\Ima(\ket{\psi})=0.005$ and $b_f=0.22$,
with $c_0=0.18$ for the catalytic strategies.
The imaginarity of the catalyst is listed as
$c_0\to c_1\to c_2$.
Here $p_1$ and $p_2$ are the conditional
success probabilities in the first and
second rounds, respectively.}
\centering
\begin{tabular*}{\textwidth}
    {@{\extracolsep{\fill}}lcccc@{}}
\noalign{\hrule height 0.8pt}
Strategy
& Catalyst imaginarity
& $p_1$
& $p_2$
& Overall success probability \\
\hline
No catalyst
& $0\to0\to0$
& $0.023$
& $0.023$
& $0.045$ \\
Full depletion
& $0.18\to0\to0$
& $0.571$
& $0.023$
& $0.581$ \\
Optimal retention
& $0.18\to0.10\to0$
& $0.377$
& $0.377$
& $0.612$ \\
\noalign{\hrule height 0.8pt}
\end{tabular*}
\end{table*}

To illustrate how the retained imaginarity
affects the overall success probability,
consider an example where $\Ima(\ket{\psi})=0.005$,
$b_f=0.22$, and the initial catalyst has
$c_0=0.18$.
Eq.~\eqref{eq:v-two-optimal-c1}
gives $c_1^\star=0.10$.
Table~\ref{tab:v-two-attempts} compares
the no catalyst, full depletion, and
optimal retention strategies.
Under full depletion, the first-round success probability is maximized at $c_1=0$,
giving $p_1=0.571$.
However, if the first round fails,
the imaginarity of the catalyst is zero
in the second round, and $p_2=0.023$.
The overall success probability is $0.581$.
Under optimal retention, $c_1^\star=0.10$
is retained after failure, giving
$p_1=p_2=0.377$.
Although the success probability of the first round
is lower, the overall success probability
increases to $0.612$.

\begin{figure*}[t]
  \centering
  \includegraphics[width=\textwidth]
  {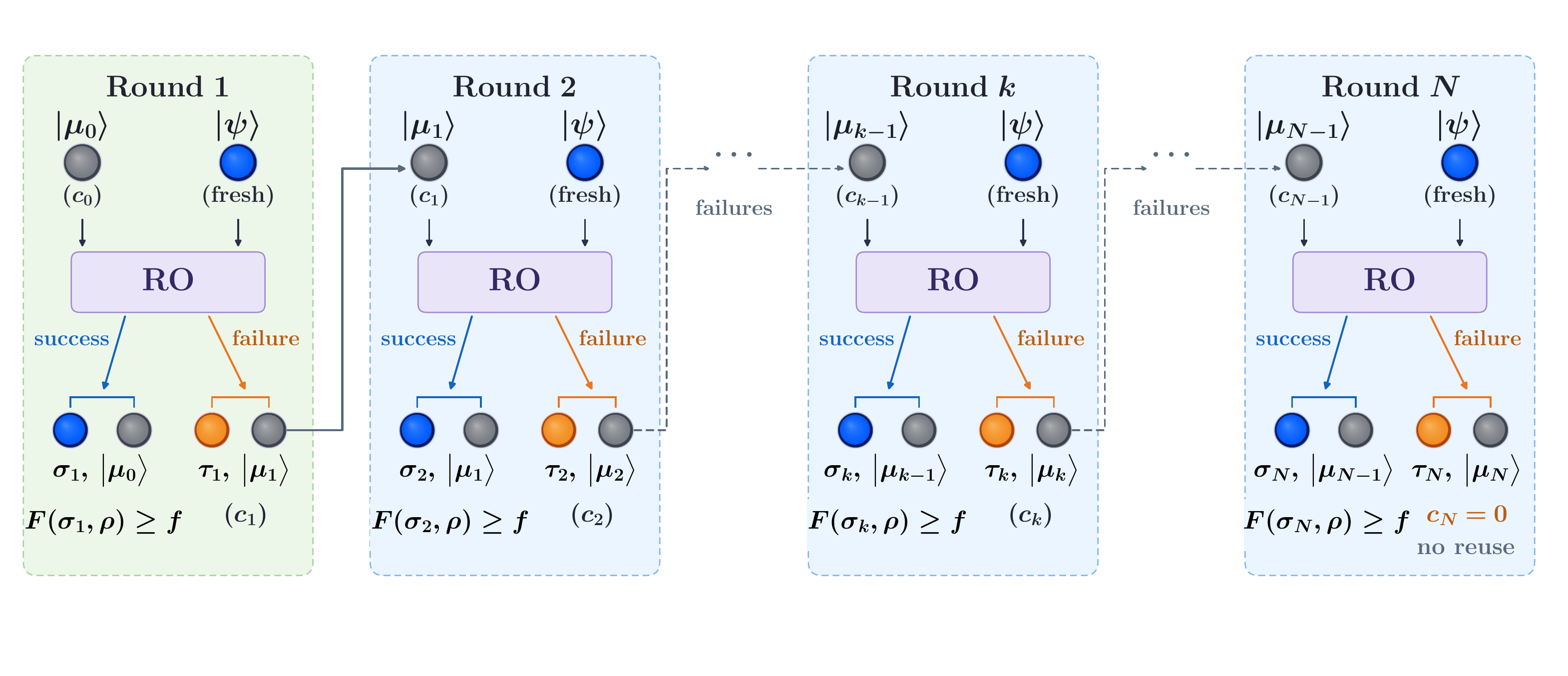}
  \caption{
  Sequential reuse of a pure catalyst over
at most $N$ rounds.
In round $k$, a fresh copy of the input
state $\ket{\psi}$ is used with the catalyst
$\ket{\mu_{k-1}}$.
On the success branch, the catalyst is
recovered exactly.
On the failure branch, it is left in the
pure state $\ket{\mu_k}$ and is available
for the next round.
Here, $c_j=\Ima(\ket{\mu_j})$.}
  \label{fig:sequential_catalysis}
\end{figure*}



For $N$ rounds, the conversion proceeds
as shown in Fig.~\ref{fig:sequential_catalysis}.
Round $k$ is performed only if all preceding
rounds have failed.
In round \(k\), the input is the joint state $\ket{\psi} \otimes \ket{\mu_{k-1}}$.
If round $k$ succeeds, the catalyst
$\ket{\mu_{k-1}}$ is returned exactly,
and no further round is performed.
If it fails, the imaginarity of the catalyst is retained in a pure state $\ket{\mu_k}$, which is used in the next round if $k<N$.
Since the catalyst is not needed after
the final round, we set $c_N=0$, giving
\begin{equation}
    c_0\geq c_1\geq\cdots\geq c_N=0.
    \label{eq:v-n-catalyst-sequence}
\end{equation}

From Eq.~\eqref{eq:v-two-failure},
the failure probability in round $k$ is
\begin{equation}
    q_k
    =
    q^\star(c_{k-1},c_k)
    =
    \frac{(1-2c_{k-1})(b-a)}
    {c_{k-1}-c_k+(1-2c_{k-1})b}.
    \label{eq:v-n-qk}
\end{equation}
Therefore, the overall success probability is
\begin{equation}
    P_N
    =
    1-\prod_{k=1}^{N}
    q^\star(c_{k-1},c_k).
    \label{eq:v-n-success}
\end{equation}

\begin{theorem}
For $N$ rounds, suppose the initial imaginarity
of the catalyst is $c_0$, with $0<c_0<1/2$.
The optimal overall success probability is 
\begin{equation}
    P_N^\star
    =
    1-
    \left[
    \frac{2(b-a)}
    {(1-2c_0)^{-1/N}-(1-2b)}
    \right]^N.
    \label{eq:v-n-optimal-probability}
\end{equation}
This maximum is attained when the imaginarity
of the catalyst after failure in round $k$ is
\begin{equation}
    c_k^\star
    =
    \frac{1-(1-2c_0)^{1-k/N}}{2},
    \qquad k=0,\ldots,N.
    \label{eq:v-n-optimal-ck}
\end{equation}
\end{theorem}

The sequence $\{c_k^\star\}_{k=0}^{N}$ depends only on
$c_0$ and $N$, and gives the same success
probability in every round that is reached.
The derivation is given in
Appendix~\ref{app:repeated-recovery-proof}.

\section{Conclusion}
\label{sec:conclusion}

We have studied catalytic state transformations in the resource theory of imaginarity using geometric imaginarity as the resource measure. Our analysis considers pure input states, arbitrary target states, and pure catalysts that are recovered exactly on the successful branch. 
For a target state with greater imaginarity
than the pure input state, a maximally
imaginary pure catalyst is necessary and
sufficient for deterministic conversion. For stochastic and stochastic-approximate
conversions, we derive the optimal catalytic success probabilities and demonstrate that nonmaximally imaginary catalysts can also provide a strict advantage.

We also focus on the evolution of the catalyst’s imaginarity on the failure branch.
For nonmaximally imaginary catalysts, the requirement of exact recovery in both branches eliminates this improvement of the optimal success probability. However, if we relax this requirement to allow the catalyst to remain pure and retain only a prescribed amount of imaginarity after failure, we obtain an analytical expression for the optimal success probability. A strict catalytic advantage requires the
imaginarity of the catalyst to decrease upon failure. This reveals a fundamental tradeoff between the success probability of the current conversion and the resources retained for later use.
We extend our analysis to $N$ rounds. In each round, a fresh
copy of the input state is used, while the
catalyst retained after failure is reused
in subsequent rounds. 
The overall success probability is maximized by optimizing the retained imaginarity after each failure. 



In this work, we assume that both input states and catalysts are pure. The optimal conversion probability remains unknown when they are mixed. Moreover, our analysis of catalytic imaginarity transformation relies on the geometric measure of imaginarity. It is worth further exploring whether similar conclusions hold for other imaginarity measures.


\begin{acknowledgments}
We thank Shuanping Du and Zhaofang Bai for valuable discussions. This work is supported by Natural Science Foundation of Henan (252300420374).
\end{acknowledgments}


\section*{Data availability}

All numerical results and figures in this article
can be reproduced from the analytical expressions
and parameter values provided in the text.
No external datasets were used.

\appendix

\section{Proof of Theorem~\ref{thm:failure_imaginarity}}
\label{app:retained_imaginarity}

The optimal success probability is obtained
by deriving an upper bound and showing that
it can be attained.
Let $a=\Ima(\ket{\psi})$ and
$\ket{\Omega}=\ket{\psi}\otimes\ket{\mu}$.
Throughout, $0\leq a<b_f\leq1/2$,
$0<c<1/2$, and $0\leq c'\leq c$.

For the output state $\sigma$, $F(\sigma,\rho)\geq f$.
It follows from the definition of $b_f$ that
$\Ima(\sigma)\geq b_f$.
Using strong monotonicity together with convexity of $\Ima$, it follows that
\begin{align}
c+(1-2c)a
&=\Ima(\ket{\Omega})
\nonumber\\
&\geq
p\Ima(\sigma\otimes\ket{\mu}\bra{\mu})
\nonumber\\
&\quad+
(1-p)\Ima(\tau\otimes\ket{\mu'}\bra{\mu'})
\nonumber\\
&\geq
p[c+(1-2c)b_f]+(1-p)c'.
\end{align}
Hence,
\begin{equation}
p\leq
\frac{c-c'+(1-2c)a}
{c-c'+(1-2c)b_f} =: p^\star.
\label{eq:retention_upper_bound}
\end{equation}

To attain this bound, the successful
system output can be chosen as a state
$\sigma_f$ satisfying
$F(\sigma_f,\rho)\geq f$ and
$\Ima(\sigma_f)=b_f$, while the system
output after failure can be a real
pure state $\ket{r}$.
The catalyst is returned in its initial state
$\ket{\mu}$ on the success branch and left
in the prescribed pure state $\ket{\mu'}$
on the failure branch.

It remains to construct a real operation
realizing these outputs with probabilities
$p^\star$ and $1-p^\star$ on two branches.
By Lemma~2 in Ref.~\cite{Kondra2023},
$\sigma_f$ admits a pure-state decomposition
\[
\sigma_f=\sum_j s_j\ket{\phi_j}\bra{\phi_j},
\qquad
\Ima(\ket{\phi_j})=b_f,
\]
where $\sum_j s_j=1$.
The corresponding ensemble of pure joint output states is
\begin{equation}
\bigl\{
(p^\star s_j,
\ket{\phi_j}\otimes\ket{\mu})
\bigr\}_j
\cup
\bigl\{
(1-p^\star,
\ket{r}\otimes\ket{\mu'})
\bigr\}.
\label{eq:success_failure_ensemble}
\end{equation}




Consider the input state
$\ket{\Omega}=\ket{\psi}\otimes\ket{\mu}$
and the ensemble in
Eq.~\eqref{eq:success_failure_ensemble}.
Since $\sum_j s_j=1$, the ensemble weights
satisfy
$\sum_j p^\star s_j+(1-p^\star)=1$.
Using $\Ima(\ket{\phi_j})=b_f$, $\Ima(\ket{r})=0$, and Eq.~\eqref{eq:retention_upper_bound}, we obtain
\begin{equation*}
\begin{split}
&\sum_j p^\star s_j
|\braket{\phi_j^*|\phi_j}|
|\braket{\mu^*|\mu}|
+(1-p^\star)
|\braket{r^*|r}|
|\braket{\mu'^*|\mu'}|\\
&=(1-2a)(1-2c)
=|\braket{\Omega^*|\Omega}|.
\end{split}
\end{equation*}
Thus, the ensemble satisfies the conditions
of Propositions~1 and~2 of
Ref.~\cite{Du2025}.
There exist real Kraus operators
$M_j$ and $M_{\rm f}$ such that
\begin{align}
M_j\ket{\Omega}
&=\sqrt{p^\star s_j}\,
\ket{\phi_j}\otimes\ket{\mu},
\nonumber\\
M_{\rm f}\ket{\Omega}
&=\sqrt{1-p^\star}\,
\ket{r}\otimes\ket{\mu'},
\end{align}
with
$\sum_j M_j^\dagger M_j+
M_{\rm f}^\dagger M_{\rm f}=\mathbb I$.

The resulting real operation,
defined by
$\mathcal E_{\rm s}(X)
=\sum_j M_jXM_j^\dagger$
and
$\mathcal E_{\rm f}(X)
=M_{\rm f}XM_{\rm f}^\dagger$,
realizes the desired ensemble
and attains $p^\star$.

\section{Proof of the Optimal Overall Success Probabilities}
\label{app:repeated-recovery-proof}

The optimal retention strategy minimizes the
probability that all rounds fail.
Let $a=\Ima(\ket{\psi})$ and $b=b_f$,
with $0\leq a<b\leq1/2$.
For $0<c_0<1/2$ and
$c_0\geq c_1\geq\cdots\geq c_N=0$,
the failure probability in
Eq.~\eqref{eq:v-n-qk} can be simplified
by introducing
\begin{equation}
    r_k=\frac{1-2c_k}{1-2c_{k-1}},
    \qquad k=1,\ldots,N.
    \label{eq:app-n-ratio}
\end{equation}
Obviously, $r_k\geq1$
and their product is
\begin{equation}
    \prod_{k=1}^{N}r_k
    =\frac{1-2c_N}{1-2c_0}
    =\frac{1}{1-2c_0},
    \label{eq:app-n-ratio-product}
\end{equation}
which is determined by $c_0$.
By Eq.~\eqref{eq:app-n-ratio}, 
$c_{k-1}-c_k
=(1-2c_{k-1})(r_k-1)/2$. Substituting it into Eq.~\eqref{eq:v-n-qk}, it follows that the failure probability in round $k$ is written as
\begin{equation}
    q_k=\frac{2(b-a)}{r_k-(1-2b)}.
    \label{eq:app-n-failure-x}
\end{equation}

\subsection{Two-Round Case}

The probability that both rounds fail is
\begin{equation}
    Q_2(c_1)=q_1q_2
    =\frac{[2(b-a)]^2}
    {[r_1-(1-2b)][r_2-(1-2b)]}.
    \label{eq:app-two-total-failure}
\end{equation}
Expanding the denominator gives
\begin{align}
&[r_1-(1-2b)][r_2-(1-2b)]
\nonumber\\
&=\frac{1}{1-2c_0}
 -(1-2b)(r_1+r_2)
\nonumber\\
&\quad +(1-2b)^2.
\label{eq:app-two-denominator}
\end{align}
For $b<1/2$, maximizing
$P_2(c_1)=1-Q_2(c_1)$ therefore
reduces to minimizing $r_1+r_2$
under $r_1r_2=(1-2c_0)^{-1}$.
Notice that  
\[
    r_1+r_2\geq2\sqrt{r_1r_2}
    =2(1-2c_0)^{-1/2},
\]
by the mean inequality,
with equality if and only if
$r_1=r_2=(1-2c_0)^{-1/2}$.
From Eq.~\eqref{eq:app-n-ratio}, it follows
\[
    c_1^\star
    =\frac{1-\sqrt{1-2c_0}}{2},
\]
which proves Eq.~\eqref{eq:v-two-optimal-c1}.

By Eq.~\eqref{eq:app-n-failure-x}, the failure probabilities
in the two rounds are
\[
    q_1^\star
    =\frac{2(b-a)}
    {(1-2c_0)^{-1/2}-(1-2b)}=q_2^\star.
\]
Hence the optimal overall success
probability is
\[
    P_2^\star
    =1-\left[
    \frac{2(b-a)}
    {(1-2c_0)^{-1/2}-(1-2b)}
    \right]^2,
\]
which proves
Eq.~\eqref{eq:v-two-optimal-probability}.

For $b=1/2$, the denominator of
Eq.~\eqref{eq:app-two-total-failure}
reduces to $r_1r_2$.
Hence,
\begin{equation}
P_2(c_1)=1-Q_2(c_1)
=1-(1-2a)^2(1-2c_0)=P_2^\star,
\end{equation}
which is independent of $c_1$.
Thus every $c_1\in[0,c_0]$ attains
the optimal overall success probability $P_2^\star$.

\subsection{$N$-Round Case}

For $N$ rounds, the corresponding
probability of failure is
\begin{equation}
    Q_N=\prod_{k=1}^{N}q_k
    =\frac{[2(b-a)]^N}
    {\displaystyle\prod_{k=1}^{N}
    [r_k-(1-2b)]}.
    \label{eq:app-n-total-failure}
\end{equation}
Maximizing
the overall success probability $P_N=1-Q_N$
amounts to maximizing the denominator
of Eq.~\eqref{eq:app-n-total-failure}
subject to
Eq.~\eqref{eq:app-n-ratio-product}.

For $b<1/2$, let
$g(t)=\log[e^t-(1-2b)]$ for $t\geq0$.
This function is strictly concave. By Eq.~\eqref{eq:app-n-ratio-product},
$\sum_{k=1}^N\log r_k
=-\log(1-2c_0)$.
From Jensen's inequality, it follows that
\begin{align}
&\sum_{k=1}^{N}
\log[r_k-(1-2b)]
\nonumber\\
&=\sum_{k=1}^{N}
g(\log r_k)
\nonumber\\
&\leq
N\times g\!\left(
\frac1N\sum_{k=1}^{N}\log r_k
\right)
\nonumber\\
&=\log\!\left[
(1-2c_0)^{-1/N}-(1-2b)
\right]^N.
\label{eq:app-n-jensen}
\end{align}
Exponentiating gives the bound
\begin{equation}
    \prod_{k=1}^{N}[r_k-(1-2b)]
    \leq
    \left[
    (1-2c_0)^{-1/N}-(1-2b)
    \right]^N.
    \label{eq:app-n-product-bound}
\end{equation}
Strict concavity implies that equality
holds if and only if
\begin{equation}
    r_1=\cdots=r_N
    =(1-2c_0)^{-1/N}.
    \label{eq:app-n-equal-ratio}
\end{equation}
Obviously, $r_k\geq1$,
and hence the optimal overall success probability can be obtained.

Combining Eqs.~\eqref{eq:app-n-ratio}
and~\eqref{eq:app-n-equal-ratio}, we obtain
\begin{equation}
c_k^\star=
\frac{1-(1-2c_0)^{1-k/N}}{2},
\qquad k=0,\ldots,N.
\end{equation}
This sequence satisfies
$c_0\geq c_1^\star\geq\cdots
\geq c_N^\star=0$.

The failure probability
in every reached round is
\[
    q_k^\star
    =\frac{2(b-a)}
    {(1-2c_0)^{-1/N}-(1-2b)}.
\]
Consequently,
\[
    P_N^\star
    =1-\left[
    \frac{2(b-a)}
    {(1-2c_0)^{-1/N}-(1-2b)}
    \right]^N,
\]
which proves
Eq.~\eqref{eq:v-n-optimal-probability}.

For $b=1/2$, the failure probability
in each round reduces to
\[
q_k=(1-2a)
\frac{1-2c_{k-1}}{1-2c_k}.
\]
Consequently,
\begin{equation}
P_N=1-(1-2a)^N(1-2c_0),
\end{equation}
independently of
$c_1,\ldots,c_{N-1}$.
Thus, every retention sequence satisfying
$c_0 \geq c_1 \geq \cdots \geq c_N = 0$
attains the optimal overall success
probability $P_N^\star$.

\bibliography{imaginarity_failure_memory_references_v12}

@article{Chitambar2019,
  author  = {Chitambar, Eric and Gour, Gilad},
  title   = {Quantum resource theories},
  journal = {Rev. Mod. Phys.},
  volume  = {91},
  pages   = {025001},
  year    = {2019},
  doi     = {10.1103/RevModPhys.91.025001}
}

@article{Brandao2015,
  author  = {Brand{\~a}o, Fernando G. S. L. and Gour, Gilad},
  title   = {Reversible Framework for Quantum Resource Theories},
  journal = {Phys. Rev. Lett.},
  volume  = {115},
  pages   = {070503},
  year    = {2015},
  doi     = {10.1103/PhysRevLett.115.070503}
}

@article{Gour2017,
  author  = {Gour, Gilad},
  title   = {Quantum resource theories in the single-shot regime},
  journal = {Phys. Rev. A},
  volume  = {95},
  pages   = {062314},
  year    = {2017},
  doi     = {10.1103/PhysRevA.95.062314}
}

@article{Liu2019,
  author  = {Liu, Zi-Wen and Bu, Kaifeng and Takagi, Ryuji},
  title   = {One-Shot Operational Quantum Resource Theory},
  journal = {Phys. Rev. Lett.},
  volume  = {123},
  pages   = {020401},
  year    = {2019},
  doi     = {10.1103/PhysRevLett.123.020401}
}

@article{Vidal1999,
  author  = {Vidal, Guifr{\'e}},
  title   = {Entanglement of Pure States for a Single Copy},
  journal = {Phys. Rev. Lett.},
  volume  = {83},
  pages   = {1046--1049},
  year    = {1999},
  doi     = {10.1103/PhysRevLett.83.1046}
}

@article{Regula2022,
  author  = {Regula, Bartosz},
  title   = {Probabilistic Transformations of Quantum Resources},
  journal = {Phys. Rev. Lett.},
  volume  = {128},
  pages   = {110505},
  year    = {2022},
  doi     = {10.1103/PhysRevLett.128.110505}
}

@article{Vidal2000,
  author  = {Vidal, Guifr{\'e} and Jonathan, Daniel and Nielsen, Michael A.},
  title   = {Approximate transformations and robust manipulation of bipartite pure-state entanglement},
  journal = {Phys. Rev. A},
  volume  = {62},
  pages   = {012304},
  year    = {2000},
  doi     = {10.1103/PhysRevA.62.012304}
}

@article{Kondra2025,
  author  = {Kondra, Tulja Varun and Datta, Chandan and Streltsov, Alexander},
  title   = {Stochastic approximate state conversion for entanglement and general quantum resource theories},
  journal = {Quantum},
  volume  = {9},
  pages   = {1929},
  year    = {2025},
  doi     = {10.22331/q-2025-12-05-1929}
}

@article{Hickey2018,
  author  = {Hickey, Alexander and Gour, Gilad},
  title   = {Quantifying the Imaginarity of Quantum Mechanics},
  journal = {J. Phys. A: Math. Theor.},
  volume  = {51},
  pages   = {414009},
  year    = {2018},
  doi     = {10.1088/1751-8121/aabe9c}
}

@article{Wu2021PRL,
  author  = {Wu, Kang-Da and Kondra, Tulja Varun and Rana, Swapan
             and Scandolo, Carlo Maria and Xiang, Guo-Yong
             and Li, Chuan-Feng and Guo, Guang-Can
             and Streltsov, Alexander},
  title   = {Operational Resource Theory of Imaginarity},
  journal = {Phys. Rev. Lett.},
  volume  = {126},
  pages   = {090401},
  year    = {2021},
  doi     = {10.1103/PhysRevLett.126.090401}
}

@article{Wu2021PRA,
  author  = {Wu, Kang-Da and Kondra, Tulja Varun and Rana, Swapan
             and Scandolo, Carlo Maria and Xiang, Guo-Yong
             and Li, Chuan-Feng and Guo, Guang-Can
             and Streltsov, Alexander},
  title   = {Resource theory of imaginarity: Quantification and state conversion},
  journal = {Phys. Rev. A},
  volume  = {103},
  pages   = {032401},
  year    = {2021},
  doi     = {10.1103/PhysRevA.103.032401}
}

@article{Kondra2023,
  author  = {Kondra, Tulja Varun and Datta, Chandan and Streltsov, Alexander},
  title   = {Real quantum operations and state transformations},
  journal = {New J. Phys.},
  volume  = {25},
  pages   = {093043},
  year    = {2023},
  doi     = {10.1088/1367-2630/acf9c4}
}

@article{Du2025,
  author  = {Du, Shuanping and Bai, Zhaofang},
  title   = {Quantifying imaginarity in terms of pure-state imaginarity},
  journal = {Phys. Rev. A},
  volume  = {111},
  pages   = {022405},
  year    = {2025},
  doi     = {10.1103/PhysRevA.111.022405}
}

@article{Guo2025,
  author  = {Guo, Meng-Li and Li, Bo and Fei, Shao-Ming},
  title   = {Geometric-like imaginarity: Quantification and state conversion},
  journal = {Sci. China Phys. Mech. Astron.},
  volume  = {68},
  pages   = {220311},
  year    = {2025},
  doi     = {10.1007/s11433-024-2514-x}
}

@article{Wu2024Distributed,
  author  = {Wu, Kang-Da and Kondra, Tulja Varun
             and Scandolo, Carlo Maria and Rana, Swapan
             and Xiang, Guo-Yong and Li, Chuan-Feng
             and Guo, Guang-Can and Streltsov, Alexander},
  title   = {Resource theory of imaginarity in distributed scenarios},
  journal = {Commun. Phys.},
  volume  = {7},
  pages   = {171},
  year    = {2024},
  doi     = {10.1038/s42005-024-01649-y}
}

@article{Fernandes2024,
  author  = {Fernandes, Carlos and Wagner, Rafael and Novo, Leonardo
             and Galv{\~a}o, Ernesto F.},
  title   = {Unitary-Invariant Witnesses of Quantum Imaginarity},
  journal = {Phys. Rev. Lett.},
  volume  = {133},
  pages   = {190201},
  year    = {2024},
  doi     = {10.1103/PhysRevLett.133.190201}
}

@article{JonathanPlenio1999,
  author  = {Jonathan, Daniel and Plenio, Martin B.},
  title   = {Entanglement-Assisted Local Manipulation of Pure Quantum States},
  journal = {Phys. Rev. Lett.},
  volume  = {83},
  pages   = {3566--3569},
  year    = {1999},
  doi     = {10.1103/PhysRevLett.83.3566}
}

@article{DaftuarKlimesh2001,
  author  = {Daftuar, Sumit and Klimesh, Matthew},
  title   = {Mathematical structure of entanglement catalysis},
  journal = {Phys. Rev. A},
  volume  = {64},
  pages   = {042314},
  year    = {2001},
  doi     = {10.1103/PhysRevA.64.042314}
}

@article{Turgut2007,
  author  = {Turgut, S.},
  title   = {Catalytic transformations for bipartite pure states},
  journal = {J. Phys. A: Math. Theor.},
  volume  = {40},
  pages   = {12185--12212},
  year    = {2007},
  doi     = {10.1088/1751-8113/40/40/012}
}

@article{LipkaBartosik2024,
  author  = {Lipka-Bartosik, Patryk and Wilming, Henrik and Ng, Nelly H. Y.},
  title   = {Catalysis in quantum information theory},
  journal = {Rev. Mod. Phys.},
  volume  = {96},
  pages   = {025005},
  year    = {2024},
  doi     = {10.1103/RevModPhys.96.025005}
}

@article{BrandaoThermo2015,
  author  = {Brand{\~a}o, Fernando G. S. L. and Horodecki, Micha{\l}
             and Ng, Nelly H. Y. and Oppenheim, Jonathan
             and Wehner, Stephanie},
  title   = {The second laws of quantum thermodynamics},
  journal = {Proc. Natl. Acad. Sci. USA},
  volume  = {112},
  pages   = {3275--3279},
  year    = {2015},
  doi     = {10.1073/pnas.1411728112}
}

@article{Ng2015,
  author  = {Ng, Nelly H. Y. and Man{\v{c}}inska, Laura and Cirstoiu, Cristina
             and Eisert, Jens and Wehner, Stephanie},
  title   = {Limits to catalysis in quantum thermodynamics},
  journal = {New J. Phys.},
  volume  = {17},
  pages   = {085004},
  year    = {2015},
  doi     = {10.1088/1367-2630/17/8/085004}
}

@article{Datta2023,
  author  = {Datta, Chandan and Kondra, Tulja Varun and Miller, Marek
             and Streltsov, Alexander},
  title   = {Catalysis of entanglement and other quantum resources},
  journal = {Rep. Prog. Phys.},
  volume  = {86},
  pages   = {116002},
  year    = {2023},
  doi     = {10.1088/1361-6633/acfbec}
}

@article{Feng2004,
  author  = {Feng, Yuan and Duan, Runyao and Ying, Mingsheng},
  title   = {When catalysis is useful for probabilistic entanglement transformation},
  journal = {Phys. Rev. A},
  volume  = {69},
  pages   = {062310},
  year    = {2004},
  doi     = {10.1103/PhysRevA.69.062310}
}

@article{Feng2005,
  author  = {Feng, Yuan and Duan, Runyao and Ying, Mingsheng},
  title   = {Catalyst-Assisted Probabilistic Entanglement Transformation},
  journal = {IEEE Trans. Inf. Theory},
  volume  = {51},
  pages   = {1090--1101},
  year    = {2005},
  doi     = {10.1109/TIT.2004.842574}
}

@article{Bu2016,
  author  = {Bu, Kaifeng and Singh, Uttam and Wu, Junde},
  title   = {Catalytic coherence transformations},
  journal = {Phys. Rev. A},
  volume  = {93},
  pages   = {042326},
  year    = {2016},
  doi     = {10.1103/PhysRevA.93.042326}
}

@article{Fang2018,
  author  = {Fang, Kun and Wang, Xin and Lami, Ludovico and Regula, Bartosz
             and Adesso, Gerardo},
  title   = {Probabilistic Distillation of Quantum Coherence},
  journal = {Phys. Rev. Lett.},
  volume  = {121},
  pages   = {070404},
  year    = {2018},
  doi     = {10.1103/PhysRevLett.121.070404}
}

@article{Liu2020,
  author  = {Liu, C. L. and Zhou, D. L.},
  title   = {Catalyst-assisted probabilistic coherence distillation for mixed states},
  journal = {Phys. Rev. A},
  volume  = {101},
  pages   = {012313},
  year    = {2020},
  doi     = {10.1103/PhysRevA.101.012313}
}

@article{ZhangLi2024,
  author  = {Zhang, Linshuai and Li, Nan},
  title   = {Can imaginarity be broadcast via real operations?},
  journal = {Commun. Theor. Phys.},
  volume  = {76},
  pages   = {115104},
  year    = {2024},
  doi     = {10.1088/1572-9494/ad6de5}
}

@article{ZhangLiLuo2024,
  author  = {Zhang, Zijian and Li, Nan and Luo, Shunlong},
  title   = {Broadcasting of imaginarity},
  journal = {Phys. Rev. A},
  volume  = {110},
  pages   = {052439},
  year    = {2024},
  doi     = {10.1103/PhysRevA.110.052439}
}

@article{FangLiu2026,
  author  = {Fang, Kun and Liu, Zi-Wen},
  title   = {One-shot distillation with constant overhead using catalysts},
  journal = {Nat. Commun.},
  volume  = {17},
  pages   = {6010},
  year    = {2026},
  doi     = {10.1038/s41467-026-72686-0}
}

@article{Jozsa1994,
  author  = {Jozsa, Richard},
  title   = {Fidelity for Mixed Quantum States},
  journal = {J. Mod. Opt.},
  volume  = {41},
  pages   = {2315--2323},
  year    = {1994},
  doi     = {10.1080/09500349414552171}
}

@article{Sharma2026,
  author  = {Sharma, Hemant and Mokeev, Aleksandr and Helsen, Jonas and Borregaard, Johannes},
  title   = {Catalytic entanglement transformations with noisy hardware},
  journal = {Quantum},
  volume  = {10},
  pages   = {2117},
  year    = {2026},
  doi     = {10.22331/q-2026-05-29-2117}
}

@article{GuoHuang2025,
  author  = {Guo, Meng-Li and Huang, Si-Yin and Li, Bo and Fei, Shao-Ming},
  title   = {Quantifying the Imaginarity via Different Distance Measures},
  journal = {Adv. Quantum Technol.},
  volume  = {8},
  pages   = {2400562},
  year    = {2025},
  doi     = {10.1002/qute.202400562}
}

@article{Shi2025,
  author  = {Shi, Xian},
  title   = {Erasing imaginarity: An operational method},
  journal = {Phys. Rev. A},
  volume  = {111},
  pages   = {L050401},
  year    = {2025},
  doi     = {10.1103/PhysRevA.111.L050401}
}

@article{Sun2025,
  author  = {Sun, Yanna and Ren, Ruonan and Wang, Yiran and Li, Yongming},
  title   = {Analysis of the relationship between imaginarity and entanglement},
  journal = {Phys. Rev. A},
  volume  = {111},
  pages   = {032425},
  year    = {2025},
  doi     = {10.1103/PhysRevA.111.032425}
}

@article{GuoGuan2026,
  author  = {Guo, Meng-Li and Guan, Nannan and Li, Bo and Hu, Bin and Fei, Shao-Ming},
  title   = {Imaginarity as a resource within quantum coherence: Geometric decomposition and operational conversion},
  journal = {Phys. Rev. A},
  volume  = {114},
  pages   = {012435},
  year    = {2026},
  doi     = {10.1103/kj4y-ttqb}
}

@article{Nakayama2026,
  author  = {Nakayama, Yasuaki and Takeuchi, Yuki and Akibue, Seiseki},
  title   = {Uniqueness of imaginarity-assisted exact transformation from real orthogonal operations to arbitrary unitary operations},
  journal = {Sci. Rep.},
  year    = {2026},
  doi     = {10.1038/s41598-026-70782-1}
}

\end{document}